\documentclass[aps,prl,twocolumn,showpacs,preprintnumbers,superscriptaddress,dblfloatfix]{revtex4-1}
\usepackage{graphicx}   
\usepackage{dcolumn}   
\usepackage{bm}        
\usepackage{amssymb}    
\usepackage{setspace}
\usepackage{amsmath, amssymb, setspace}
\usepackage{array}
\usepackage{booktabs}
\usepackage{mathrsfs}
\usepackage{indentfirst}
\usepackage{slashed}
\usepackage{float}
\usepackage{lmodern}
\usepackage{hyperref}
\usepackage{xcolor}
\usepackage{multirow}
\usepackage{epstopdf}
\usepackage{ulem}
\usepackage{tabularx,lipsum}
\usepackage[justification=raggedright]{caption}
\usepackage[flushleft]{threeparttable}
\usepackage{hyperref}
\usepackage{soul}

\begin{document}

\title{Primordial Black Holes and $r$-Process Nucleosynthesis}

\preprint
{IPMU17-0048}

\author{George M. Fuller}
\email[]{gfuller@ucsd.edu}
\affiliation{Department of Physics, University of California, San Diego\\
La Jolla, California 92093-0424, USA}

\author{Alexander Kusenko}
\email[]{kusenko@ucla.edu}
\affiliation{Department of Physics and Astronomy, University of California, Los Angeles\\
Los Angeles, CA 90095-1547, USA}
\affiliation{Kavli Institute for the Physics and Mathematics of the Universe (WPI), UTIAS\\
The University of Tokyo, Kashiwa, Chiba 277-8583, Japan}

\author{Volodymyr Takhistov}
\email[]{vtakhist@physics.ucla.edu}
\affiliation{Department of Physics and Astronomy, University of California, Los Angeles\\
Los Angeles, CA 90095-1547, USA}

\date{\today}

\begin{abstract}
We show that some or all of the inventory of $r$-process nucleosynthesis can be produced in interactions of primordial black holes (PBHs) with neutron stars (NSs) if PBHs with masses ${10}^{-14}\,{\rm M}_\odot < {\rm M}_{\rm PBH} < {10}^{-8}\,{\rm M}_\odot$ make up a few percent or more of the dark matter. A PBH captured by a neutron star (NS) sinks to the center of the NS and consumes it from the inside. When this occurs in a rotating millisecond-period NS, the resulting spin-up ejects $\sim 0.1-0.5\,{\rm M}_{\odot}$ of relatively cold neutron-rich material. This ejection process and the accompanying decompression and decay of nuclear matter can produce electromagnetic transients, such as a kilonova-type afterglow and fast radio bursts. These transients are not accompanied by significant gravitational radiation or neutrinos, allowing such events to be differentiated from compact object mergers occurring within the distance sensitivity limits of gravitational wave observatories. 
The PBH-NS destruction scenario is consistent with pulsar and NS statistics, the dark matter content and spatial distributions in the Galaxy and Ultra Faint Dwarfs (UFD), as well as with the $r$-process content and evolution histories in these sites. Ejected matter is heated by beta decay, which leads to emission of positrons in an amount consistent with the observed 511-keV line from the Galactic Center.
\end{abstract}

\maketitle

Primordial black holes (PBHs) can account for all or part of the dark matter (DM)~\cite{Zeldovich:1967,Hawking:1971ei,Carr:1974nx,GarciaBellido:1996qt,Khlopov:2008qy,Frampton:2010sw,Kawasaki:2012kn,Kawasaki:2016pql,Cotner:2016cvr,Carr:2016drx,Inomata:2016rbd,Inomata:2017okj,Georg:2017mqk}.  If a PBH is captured by a neutron star (NS), it settles into the center and grows until the supply of nuclear matter is exhausted by accretion and ejection. 

In this {\it Letter} we show that NS disruptions by PBHs in DM-rich environments, such as the Galactic center (GC) and dwarf spheroidal galaxies, provide a viable site for r-process nucleosynthesis, thus offering a solution to a long-standing puzzle~\cite{Burbidge:1957vc,Arnould:2007gh,Cameron:1957,Truran:2002nj,Qian:2007vq}.~The transients accompanying NS disruption events and the positrons produced in these events are consistent with present observations, and they offer a way of testing the NS--PBH scenario in the future.

We will demonstrate that, when a PBH accretes matter inside a rapidly rotating millisecond pulsar (MSP), the resulting pulsar spin-up causes $\sim 0.1-0.5\,M_{\odot}$ of neutron-rich material to be ejected without significant heating and only modest neutrino emission. This provides a favorable setting for $r$-process nucleosynthesis,
occuring on the Galactic time-scales, which can evade several problems that have challenged the leading proposed $r$-process production sites, such as neutrino-heated winds from core collapse supernovae or binary compact object mergers (COM)~\cite{Argast:2003he,Freiburghaus:1999}. The unusual distribution of $r$-process abundances within the ultra-faint dwarf spheroidal galaxies~\cite{Ji:2016,Beniamini:2016rnw} (UFDs) is naturally explained by the rates of PBH capture in these systems. 
The rates are also consistent with the paucity of pulsars in the GC~\cite{Dexter:2013xga}.
A similar distribution of $r$-process material in UFDs can be expected from NS disruptions due to black holes produced in the NS interiors by accretion of particle dark matter onto the NS~\cite{Bramante:2016mzo}, although the rates and the implications for dark matter properties are, of course, different. The probability of PBH capture depends on both the PBH and the NS densities.  DM-rich environments, such as the GC and dwarf spheroidal galaxies, are not known to host NSs. An exception is the young magnetar~\cite{Mori:2013yda,Kennea:2013dfa}, whose age is small compared to the time scales of PBH capture.  NSs are found in the disk and the halo, as well as in the globular clusters, where the dark matter density~\cite{Bradford:2011aq,Ibata:2012eq} is too low to cause a substantial decrease in the pulsar population.  
Positrons emitted from the heated neutron-rich ejecta can account for the observed 511-keV line from the GC~\cite{Prantzos:2010wi,Lingenfelter:2009kx}. The final stages of neutron star demise can be the origin~\cite{Fuller:2014rza,Shand:2015uda} of some of the recently observed~\cite{Lorimer:2007qn} fast radio bursts (FRBs), as well as X-ray and $\gamma$-ray transients. A kilonova-type~\cite{Kasen:2014toa,Li:1998bw,Metzger:2010sy,Roberts:2011xz,Hotokezaka:2015eja,Piran:2012wd,Barnes:2013wka,Martin:2015hxa} afterglow can accompany the decompressing nuclear matter ejecta, but unlike COM, these events are not associated with a significant release of neutrinos or gravitational radiation. Therefore, future observations of gravitational waves and kilonovae will be able to distinguish between $r$-process scenarios.

\textit{Millisecond pulsars} are responsible for the predominant contribution to the nucleosynthesis initiated by PBH-induced  centrifugal ejection of neutron-rich material, since MSPs have the highest angular velocities at the time of PBH capture. The most prominent sites of $r$-process production must have a high density of MSPs as well as PBHs. The latter trace the DM spatial distribution.  The DM density is high in the GC and in the UFDs. On the other hand, the MSP density is high in molecular clouds, including the Central Molecular Zone (CMZ), and the globular clusters.  While the CMZ is located within the GC with an extremely high DM density, observations imply that the DM content of globular clusters is fairly low~\cite{Bradford:2011aq,Ibata:2012eq}. The product of the DM density and the MSP density is still sufficient to allow for some $r$-process nucleosynthesis in both the UFDs and the globular clusters, but we estimate that CMZ accounts for 10\% to 50\% of the total Galactic production.  
We include contributions from the GC (CMZ) and the rest of the halo (which may be comparable, within uncertainties)~\cite{Read:2009iv,Read:2008fh}. 

The CMZ has an approximate size of $\sim200$ pc and is located near the GC, where supernova rates are the highest. Since the DM density peaks at the GC, the CMZ is a site of frequent PBH-MSP interactions.  The pulsar formation rate~\cite{OLeary:2016cwz} in the CMZ is $1.5 \times 10^{-3}$yr$^{-1}$ ($7\%$ of the Galactic formation rate), consistent with the GeV $\gamma$-ray flux observed~\cite{TheFermi-LAT:2015kwa} from the GC by the \textit{Fermi} Large Area Telescope.  
Hence, we expect $N_p^{\text{GC}} \approx 1.5 \times 10^{7}$ NSs to be produced at the GC during the lifetime of the Galaxy, $t_G \sim10^{10}$ yr.  Roughly, 30\%-50\% of these NSs become MSPs~\cite{Lorimer:2012hy}, and the number of MSPs with a particular rotation period can then be estimated from a  population model~\cite{Cordes:1997my,FaucherGiguere:2005ny,Lorimer:2012hy}. 

Simulations and observations of UFDs imply~\cite{Ji:2016} that  $\sim 2000$ core-collapse supernovae have occurred in 10 UFDs during  $t_{\text{UFD}} \sim5\times10^8$ yr. Hence, we expect that $N_p^{\text{UFD}} \sim 10^2$ pulsars have been produced in each of these systems, on average, and we estimate the fraction of fast-rotating MSPs using a population model~\cite{Cordes:1997my}.

\textit{The black hole capture rate} can be calculated using the initial PBH mass $(m_{\rm PBH})$,  DM density and velocity dispersion (assuming a Maxwellian distribution). With the base Milky Way (MW) and UFD capture rates denoted as $F_0^{\rm MW}$ and $F_0^{\rm UFD}$, one obtains~\cite{Capela:2013yf} the full PBH-NS capture rates $F=(\Omega_{\rm PBH}/\Omega_{\rm DM} ) F_0^{\rm MW}$ and $F= (\Omega_{\rm PBH}/\Omega_{\rm DM} )F_0^{\rm UFD}$ in the MW and UFD, respectively. 
 The number of PBHs captured by a NS is $F \cdot t$, where the time $t$ is either $t_G$ or $t_{\text{UFD}}$ for the MW and the UFD, respectively.
For our analysis we consider a typical NS with mass $M_{NS} = 1.5 M_{\odot}$ and radius $R_{NS} = 10$ km.
The base NS-PBH capture rate $F_0$ is given by ~\cite{Capela:2013yf}
\begin{equation} \label{eq:f0calc}
F_0 = \sqrt{6 \pi} \dfrac{\rho_{\text{DM}}}{m_{\rm PBH}} \Big[\dfrac{R_{\rm NS} R_{\rm S}}{\overline{v}(1 - R_{\rm S}/R_{\rm NS})}\Big] \Big(1 - e^{- 3 E_{\rm loss}/(m_{\rm PBH}\overline{v}^2)}\Big)~,
\end{equation}
where $\overline{v}$ is the DM velocity dispersion, $R_{\rm S} = 2 G M_{\rm NS}$ and $R_{\rm NS}$ are the BH Schwarzschild radius and the NS radius, respectively. $E_{\rm loss}$ is the energy loss associated with PBH-NS interaction. BH capture can occur when $E_{\rm loss} > m_{\rm PBH} v_0^2/2$, with $v_0$ being the asymptotic velocity of the PBH.
Taking a uniform flux of PBHs across the star, the average energy loss for a typical NS is found to be $E_{\rm loss} \simeq 58.8  G^2 m_{\rm PBH}^2 M_{\rm NS}/R_{\rm NS}^2$. Since MSPs originate from binaries, a higher binary gravitational potential causes an increase in the capture rate. We, therefore, assume that the capture rate for MSPs is a factor 2 higher than for isolated NSs.  For example, for typical values of parameters and $m_{\rm PBH} = 10^{19}$~g, one obtains
\begin{equation}
 \left\{
                \begin{array}{ll}
                  F_0^{\rm MW} = 1.5 \times 10^{-11} {\rm /yr}\\
                  F_0^{\rm UFD} = 6.0 \times 10^{-10} {\rm /yr}~.\\
                \end{array}
              \right.
\end{equation}
For the MW we have used a velocity dispersion of 48 km/s and 105 km/s for NS and DM, respectively, as well as DM density $8.8 \times 10^2$ GeV/cm$^3$. The pulsar and DM velocity dispersions are simultaneously taken into account for the MW as described below. For UFD we have used DM velocity dispersion 2.5 km/s and DM density 10 GeV/cm$^{3}$.

A PBH could also be captured by a NS progenitor prior to supernova core collapse~\cite{Capela:2014ita}, but this does not increase the capture rates significantly.

Natal pulsar kicks can enable pulsars to escape from the region of interest. We include this effect in our calculations (see Supplemental Material~\cite{suppmat}).

\textit{Pulsar lifetimes} in the presence of PBHs with a given number density can be estimated as $\langle t_{\rm NS} \rangle = 1/F + t_{\rm loss} + t_{\rm con}$, where the first term describes the mean BH capture time, $t_{\rm loss}$ is the time for the PBH to be brought within the NS once it is gravitationally captured, and $t_{\rm con}$ is the time for the black hole to consume the NS.
For a typical NS one finds~\cite{Capela:2013yf} that $t_{\rm loss} \simeq 4.1 \times 10^4\, (m_{\rm PBH}/10^{22}\,\text{g})^{-3/2} \, {\rm yr}$. The spherical accretion rate of NS matter onto the PBH is described by the Bondi equation $d m_{\rm BH}/dt = 4 \pi \lambda_s G^2 m_{\rm BH}^2 \rho_c / v_s^3 = C_0 m_{\rm BH}^2$, where $m_{\rm BH}(t)$ is the time-dependent mass of the central black hole, $v_s$ is the sound speed, $\rho_c$ is the central density and $\lambda_s$ is a density profile parameter.
For typical NS values of~\cite{Shapiro:1983du} $v_s = 0.17, \rho_c = 10^{15}$ g/cm$^{3}$ and $\lambda_s = 0.707$ (for a star described by an $n = 3$ polytrope) we obtain that $t_{\rm con} = 10~(10^{19}~\text{g}/m_{\rm PBH})$~yr. If PBHs make up all of the DM, 
we calculate that $\langle t_{\rm NS} \rangle < 10^{12}$ yr for $10^{17}~\text{g} < m_{\rm PBH} < 10^{25}\,\text{g}$, implying that a $\mathcal{O}(1-10)\%$ fraction of pulsars should have been consumed in the age of the Galaxy.
This is consistent with observations~\cite{Dexter:2013xga} suggesting an under abundance of MSPs near the central Galactic black hole, Sgr~A$^{\ast}$. A recently discovered young, $4\times 10^4$ yr old, magnetar J1745-2900 located just 0.1 pc from the GC~\cite{Mori:2013yda,Kennea:2013dfa} is also consistent with our results, since this magnetar's age is shorter than $\langle t_{\rm NS} \rangle$. The unusual surface temperature~\cite{Zelati:2015vya} and X-ray luminosity of J1745-2900 warrants scrutiny, as this activity might be consistent with PBH destruction in progress.

\textit{Angular momentum transfer} determines the dynamics of a NS spin-up.~As the captured PBH starts to grow and consume the spinning pulsar from the inside, the radius of the neutron star decreases and angular momentum conservation forces a spin-up.  As the star contracts, the fractional change in radius is greater for accreted matter in the inner regions than it is for material further out. This could lead to differential rotation.  However, if angular momentum can be efficiently transferred outward, the star can maintain rigid-body rotation. Viscosity~\cite{Kouvaris:2013kra,Markovic:1994bu} and magnetic stresses~\cite{Markovic:1994bu} can prevent differential rotation from developing. It can be shown (see Supplemental Material ~\cite{suppmat}) that angular momentum is transferred efficiently on the relevant time scales and that Bondi accretion proceeds nearly uninterrupted throughout the BH evolution.

\textit{Ejected mass} originates from the  star spin-up when matter at the equator exceeds the escape velocity.
Using polytropic NS density runs with different indices~\cite{Shapiro:1983du}, we have calculated analytically (see end of ``Ejected Mass'' section in Supplemental Material~\cite{suppmat}) the amount of ejected material.
The results are shown in Figure~\ref{fig:ejmass} for NS victims with a range of initial rotation periods. Based on our estimates discussed in Supplemental Materials, NS with periods of a few ms can eject more than $10^{-1} M_{\odot}$ of material. A detailed calculation taking into account general relativistic effects \cite{Andersson:2000mf,Owen:1998xg} is needed to improve understanding of the ejected mass.

The number of MSPs in the disk with periods greater than $P$ is described by a power-law distributed population model~\cite{Cordes:1997my}.  We assume that the distribution in the CMZ is the same, and we normalize the total to the number of neutron stars produced in supernova explosions.  According to the population model~\cite{Cordes:1997my}, $N_{\rm MSP} \simeq 1.6 \times 10^4  \Big(1.56\, {\rm ms}/P\Big).$ Using the differential distribution $d(N_{\rm MSP})/dP$, we obtain the population-averaged ejected mass: 
\begin{equation}
\langle M_{\rm ej} \rangle = \dfrac{\int_{P_{\min}}^{\infty} \Big(\dfrac{dN_{\rm MSP}}{dP}\Big) M_{\rm ej}(P) dP}{\int_{P_{\min}}^{\infty} \Big(\dfrac{dN_{\rm MSP}}{dP}\Big)dP}~,
\end{equation}
where $P_{\min}$ is the minimal MSP period in the population, and $M_{\rm ej}(P)$ is the  ejected mass function interpolated from the distribution shown in Figure~\ref{fig:ejmass}.
 We find that the population-averaged ejected mass is $\langle M_{\rm ej} \rangle = 0.18 M_{\odot}$ and $0.1 M_{\odot}$ if we take the shortest period to be $P_{\min} = 0.7$~ms (theoretically predicted) and $P_{\min} = 1.56$~ms (observed), respectively. Since realistic nuclear matter equations of state suggest flatter NS density profiles than our polytropic approximations, our estimate is conservative and $\langle M_{\rm ej}\rangle$ can be up to a factor of few larger.  Alternative population models \cite{Lorimer:2012hy}, such as those based on \cite{FaucherGiguere:2005ny}, do not significantly alter the results.

\begin{figure}[tb]
\centering
\hspace{-1em}
\includegraphics[trim = 0.0mm 17.5mm 32.0mm .0mm, clip,width = 3.3in]{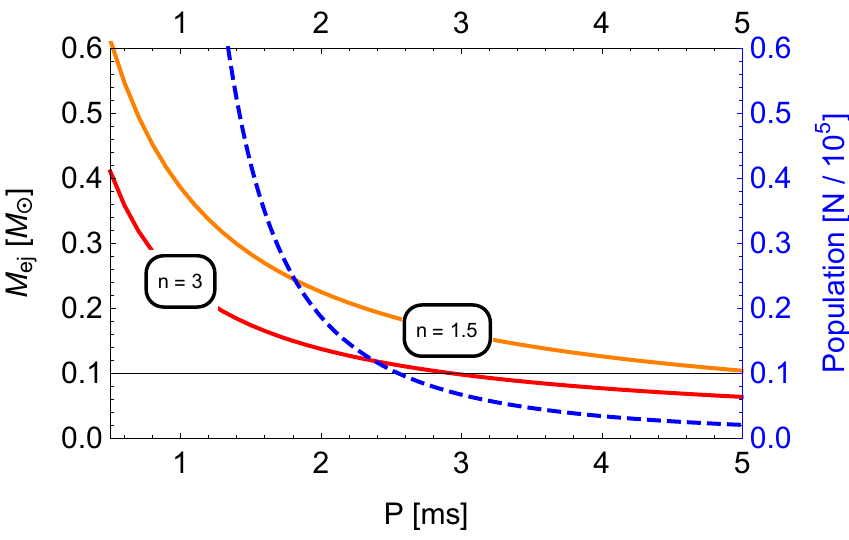}\\
\caption{
Total ejected mass ($M_{\rm ej}$) from a MSP with initial rotation period $P$ disrupted by a PBH. $n = 3$ polytrope (red) and $n = 1.5$ polytrope (orange) NS density profiles are shown. The black line indicates ejection of $0.1~M_{\odot}$. The MSP period--population distribution~\cite{Cordes:1997my} is displayed with a dashed blue line.}
\label{fig:ejmass}
\end{figure}

\textit{Nucleosynthesis}
takes place in the ejecta.
Heating accompanying the growth of a BH inside a NS results in a temperature increase near the event horizon that is only a factor of few higher than the NS surface temperature~\cite{Kouvaris:2013kra}: $T_{\rm h}/T_{\rm surf} \sim 3$. Consequently, neutrino emission is negligible and ejected material does not suffer significant heating or exposure to neutrinos.   

Decompression of the centrifugally-ejected, relatively low entropy and very low electron fraction nuclear matter in this scenario could be expected to result in a significant mass fraction of this material participating in $r$-process nucleosynthesis~\cite{Lattimer:1977,Meyer:1989,Lippuner:2015gwa,Jaikumar:2006qx,Goriely:2011vg,Gorielyproc:2016,Eichler:2014kma}.~The large neutron excess in this scenario, relatively unmolested by neutrino charged current capture-induced reprocessing of the neutron-to-proton ratio, could lead to fission cycling~\cite{Eichler:2014kma,Mumpower:2016ssd}, thereby tying together the nuclear mass number $A= 130$ and $A=195$ $r$-process abundance peaks. Unlike COM $r$-process ejecta, which will have a wide range of neutrino exposures, entropy, and electron fraction and thereby can reproduce the solar system $r$-process abundance pattern~\cite{Wanajo:2014wha}, the PBH scenario may be challenged in producing the low mass, $A < 100$, $r$-process material. 

The material ejected in the PBH-NS destruction process is heated by beta decay and fission, resulting in thermodynamic conditions and abundances closely akin to those in the COM-induced ``tidal tail" nuclear matter decompression that gives rise to kilonova-like electromagnetic signatures~\cite{Kasen:2014toa,Li:1998bw,Metzger:2010sy,Roberts:2011xz,Hotokezaka:2015eja,Piran:2012wd,Barnes:2013wka,Martin:2015hxa}. This could be a more luminous and longer duration transient compared to the classic COM-generated kilonovae, as the ejecta in the PBH scenario can have more mass than the tidal tails of COM.   

The total amount of ejected $r$-process material in the PBH-NS destruction process can be estimated via 
$M_{\rm tot}^r =  F t N_{\rm MSP} \langle M_{\rm ej} \rangle$, assuming that the bulk of the ejecta undergoes $r$-process nucleosynthesis. The overall mass of $r$-process material in the Galaxy is $M_{\rm tot}^{r, {\rm MW}}\sim {10}^4\,{\rm M}_\odot$. 
The required fraction of dark matter in the form of PBHs is 
$(\Omega_{\rm PBH}/\Omega_{DM}) = M_{\rm tot}^{r,{\rm MW}}/(F_0^{{\rm MW}} t_G N_{\rm MSP}^{{\rm GC}}\langle M_{\rm ej} \rangle)$.
If the mass of ejected r-process material in a single event is 
$0.1 - 0.5\, {\rm M}_\odot$, the PBH capture rate $ {10}^{-5} - {10}^{-6}\,{\rm Mpc}^{-3}\,{\rm yr}^{-1}$ can account for {\it all} of the $r$-process in the Galaxy.  At this rate, $10^5$ NS disruption events have occurred in the lifetime of the Galaxy.

This rate of NS disruptions in UFDs is also consistent with the observationally inferred UFD $r$-process content and with the uneven distribution of this material among the observed UFD. Observations imply that one in ten of UFDs have been a host to r-process nucleosynthesis events, which must, therefore, be rare~\cite{Ji:2016,Hirai:2015npa,Beniamini:2016rnw}.  The rate $F_0^{\rm UFD}$ implies that the probability of a NS disruption in a single UFD is about 0.1, which explains the uneven distribution. The amount of $r$-process material supplied by a single event, $\sim 0.1\, M_\odot$, is more than sufficient to explain the observations~\cite{Ji:2016,Hirai:2015npa,Beniamini:2016rnw}.  Only a small fraction $\sim (\bar{v}^{\rm UFD}/v_{\infty})\sim 10^{-3} $ of the produced r-process material is likely to remain in the shallow gravitational potential well of a UFD because it is produced with a velocity $v_\infty\sim 0.1 \, v_{\rm esc} $.  The observations are consistent with this: the required $10^{-4}M_\odot$ of r-process material is consistent with a 0.1\% fraction of the $0.1M_\odot$ produced in a single event.

We have separately fit to the $r$-process abundances for the MW and UFD, accounting for uncertainties in various quantities as described below. The combined requirements result in the allowed region of parameter space shown on Figure~\ref{fig:PBHdm}, along with the current constraints for PBH contribution to the DM abundance. The region denoted ``all $r$-process''
shows parameter space for which $r$-process observations are fully explained simultaneously in the MW and in UFDs. For our fit we have varied the input parameters over a broad range, covering significant parameter space (see Supplemental Material ~\cite{suppmat}). The enclosed region above the line can be interpreted as a constraint of $r$-process material over-production from PBH-NS interactions, subject to large uncertainties in astrophysical input parameters. Energy losses and capture rates for black holes with masses below $\sim10^{18}$ g are not well understood, and there is an uncertainty in the range of parameters for small masses.

We note that COM-produced $r$-process, with an event rate of $ {10}^{-4} - {10}^{-5}~{\rm Mpc}^{-3}\,{\rm yr}^{-1}$, could also be consistent with this analysis~\cite{Goriely:2011er}.~However, COM simulations suggest an ejecta mass of $\sim 0.01\,{\rm M}_\odot$. This would imply a COM rate near the upper end of the allowed range, if COM are to explain all of the $r$-process. Such a rate is still marginally consistent with the current Advanced LIGO (aLIGO) limits, but readily verifiable or refutable when aLIGO reaches its design sensitivity~\cite{Cote:2016vla} in a few years. Sensitivity similar to aLIGO is expected in the upcoming Advanced Virgo~\cite{TheVirgo:2014hva} (aVirgo) and KAGRA~\cite{Aso:2013eba} experiments.
A recent analysis of kilonova~\cite{Jin:2016pnm} also exhibits tension with observations and highlights the need for an extremely efficient ejection of $r$-process material in COM scenario.

\begin{figure}[tb]
\centering
\hspace{-2em}
\includegraphics[trim = 0.0mm .0mm .0mm .0mm, clip,width = 3.3in]{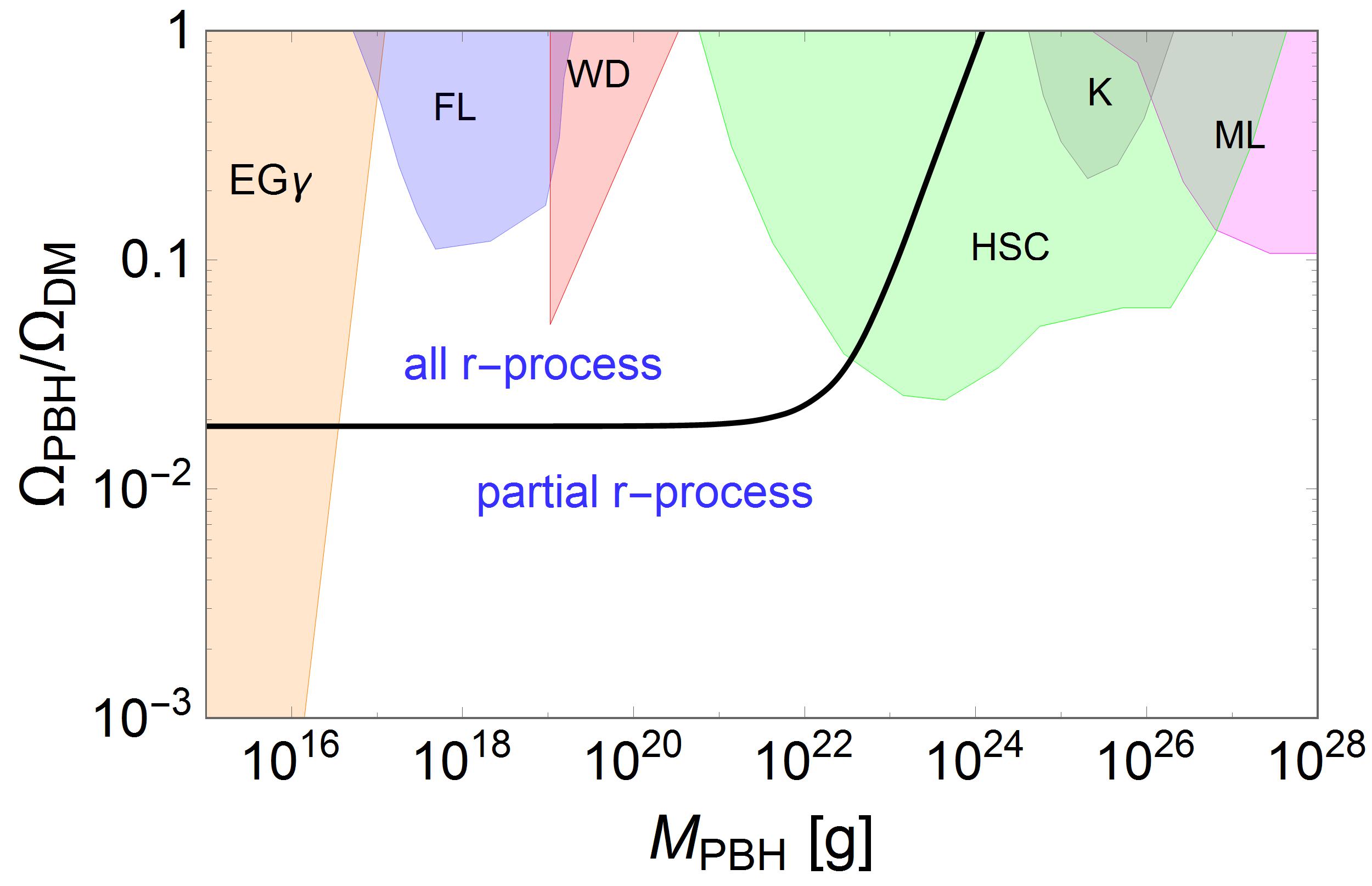}\\
\caption{
Parameter space where PBHs can account for all or partial $r$-process element production in the Milky Way and the UFDs simultaneously. 
Constraints from extragalactic $\gamma$-rays from BH  evaporation~\cite{Carr:2009jm} (EG$\gamma$), femto-lensing~\cite{Barnacka:2012bm} (FL), white dwarf abundance~\cite{Graham:2015apa} (WD), Kepler star milli/micro- lensing~\cite{Griest:2013aaa} (K), Subaru HSC micro-lensing~\cite{Niikura:2017zjd} (HSC) and MACHO/EROS/ OGLE micro-lensing~\cite{Tisserand:2006zx} (ML) are displayed.}
\label{fig:PBHdm}
\end{figure}

\textit{Positron emission}
from ejecta can explain
the observed 511~keV emission line from the Galactic central
region~\cite{Jean:2003ci}, which is consistent with the $e^+e^-$ annihilation line via positronium formation. The origin of the positrons remains unknown~\cite{Prantzos:2010wi}. The 511 keV line flux
in the bulge component is~\cite{Knodlseder:2003sv}
$\sim 10^{-3}$ photons cm$^{-2}$ s$^{-1}$.  The line can be explained through electron--positron annihilations that occur at a rate of
$\Gamma(e^+e^-\rightarrow \gamma\gamma) \sim 
 10^{50}\, {\rm yr}^{-1} $. 
Ejected cold nuclear matter expands on a dynamical time scale of
$\tau_e\sim \alpha /\sqrt{G_N\rho}=446 \alpha (\rho/{\rm g\, cm}^{-3})^{-1/2} {\rm s}~,
$
where $\alpha=0.01-10$ is a model-dependent parameter~\cite{Lattimer:1977,Meyer:1989,Goriely:2011er}. 
At the same time, beta decays and fission raise the temperature to $T \sim 0.1$~MeV~\cite{Lattimer:1977,Meyer:1989,Goriely:2011er}.  This temperature is high enough to generate a sizable equilibrium density of positrons, which leak through the surface of each clump. Taking the radius of each clump as $R\sim 0.1$~km and the density as $\rho\sim 10^8$g/cm$^3$, the total surface area of $0.1 \, M_\odot$ of ejected material is $A\sim 4\pi R^2 ( 0.1 M_\odot/\rho)/(4\pi R^3/3)\sim 10^{20} {\rm cm}^2$.  The number of positrons emitted in a single event, during the time $\tau_e$, while the temperature $T\sim 0.1$~MeV is maintained, can be estimated as $N_{e^+}\sim Av\tau_e \times 2 (m_e T/2\pi)^{3/2} \exp (-m_e/T) $, where $v$ is the average speed of positrons emitted with a relativistic $\gamma$ factor $\gamma \sim (3T/m_e)$.  If the neutron star disruption events occur in the GC on the time scale of $\tau_{d} \sim 10^5$ yr, the average rate of positron production is
\begin{equation}
R_{e^+}= N_{e^+}/\tau_d \sim 10^{50}\, {\rm yr}^{-1}~.
\end{equation}
Since the average positron energy $E_{e^+}\approx 3 T$ is below 3~MeV, the positrons do not annihilate in flight in the interstellar medium~\cite{Beacom:2005qv}. 

\textit{Fast radio bursts, kilonovae and other signatures} are expected from the PBH capture-induced NS demise.
During the final stages of the event, described by dynamical time scales of the order of a few to tens of milliseconds, $10^{41}-10^{43}$~ergs of energy  stored in the magnetic field are released.  Inside the cold NS, at temperatures below 0.4 MeV, the nuclear matter is a Type II superconductor and magnetic field is concentrated in flux tubes. A consequence of the rapid rearrangement of nuclear matter accompanying ejection is a prodigious release of electromagnetic radiation from magnetic field reconnection and decay. The resulting bursts of radio waves~\cite{Fuller:2014rza,Shand:2015uda} with duration of a few milliseconds can account for some of the observed~\cite{Lorimer:2007qn} FRBs.  One FRB is known to be a repeater, while the others appear to be one-time events. The FRB energy of $10^{41}$~erg is consistent with observations~\cite{Dolag:2014bca}. If 1--10\% of the magnetic field energy is converted to radio waves, an FRB could accompany an NS destruction event.  The rapidly evolving magnetic field can also accelerate charged particles leading to X-ray and $\gamma$-ray emission.

Detection of an ``orphaned" kilonova (macronova) within the aLIGO, aVirgo and KAGRA sensitivity distance ($\sim 200\,{\rm Mpc}$) that is {\it not} accompanied by a binary compact object in-spiral gravitational wave signal or a short $\gamma$-ray burst, but possibly associated with an FRB, would constitute an indirect argument that NS disruptions via PBH capture occur and could account for a significant fraction of the $r$-process. Sophisticated  numerical simulations of PBH-induced NS collapse and of the accompanying nucleosynthesis and electromagnetic emission (including FRB) could help enable feasible observational search strategies. The search can be further assisted by detailed mapping of chemical abundances that will be made possible by the future Hitomi-2 detector. The stakes are high, as finding evidence for PBH-NS destruction could have profound implications for our understanding of the origin of the heavy elements and for the source and composition of dark matter.     

\textit{Acknowledgments.}
We thank B.~Carr, Y.~Inoue, S.~Nagataki, R.~Rothschild, and E.~Wright for helpful discussions. 
The work of G.M.F. was supported in part by National Science Foundation Grants No. PHY-1307372 and PHY-1614864.  
The work of A.K. and V.T. was supported by the U.S. Department of Energy Grant No. DE-SC0009937. A.K. was also supported by the World Premier International Research Center Initiative (WPI), MEXT, Japan.  \(\)

\bibliography{pbhnsbib}

\newpage
\begin{titlepage}
\begin{center}
{\large{\bf Supplemental Material}}
\end{center}
\vspace{3em}
\end{titlepage}

\section{Natal Pulsar Kicks}

Only $\sim2\%$ of NS from UFDs remain in the central region on the relevant time scales~\cite{Bramante:2016mzo}. On the other hand, since some MSPs reside in binaries, their kick velocities are retarded by gravitational binding to the companion and, consequently, the number of resident neutron stars can be higher, $\sim 10\%$. We modify
the number $N_p^{\rm UFD}$ accordingly. 

To account for the pulsar velocity dispersion in the MW, we replace the star capture rate $F_0^{\rm MW}$ for stationary NS with an effective rate, combining the Maxwellian velocity distributions of DM and pulsars with a pulsar velocity dispersion~\cite{Cordes:1997my} of $\sim 48$ km s$^{-1}$.
Including the velocity dispersions of PBHs and pulsars, the modified rate is
\begin{equation}
F_0^{\rm MW} = \int d^3v_{n} f^{3D}_n (\vec{v}_n, \overline{v}_n) \int d^3v_{d} f^{3D}_d (\vec{v}_d, \overline{v}_d) F_0 (|\vec{v}_n - \vec{v}_d|)~,
\end{equation}
where the subscripts $n, d$ refer to NS and DM/PBH, respectively, while $f^{3D} (\vec{v}, \overline{v})$ denotes the appropriate 3-D Maxwellian velocity ($\vec{v}$) distribution with a mean of $\overline{v}$. The 3-D integrals are decomposed by switching to spherical coordinates, where $d^3 v_n = v_n^2 dv_n d (\cos \chi) d\nu$,
$d^3 v_d = v_d^2 dv_d d (\cos \eta) d\beta$,
with the $d(\cos\chi), d(\cos\eta)$ integrals evaluated on the interval $[-1, 1]$ and $d\beta, d\nu$ on $[0, 2 \pi]$. The vector difference of velocities decomposes as $
|\vec{v}_n - \vec{v}_d| = (v_n^2 + v_d^2 - 2 v_n v_d \cos \chi)^{1/2}$, where $\chi$ is the lab frame angle between the two velocity vectors. The velocity integrand is then weighted by the 1-D Maxwellian distribution  
\begin{equation}
f^{1D}(v,\overline{v}) = \Big(\dfrac{3}{2 \pi \overline{v}^2}\Big)^{3/2} e^{- 3 v^2/(2\overline{v}^2)}
\end{equation}
and is evaluated over the interval $[0, \infty)$.

\section{Viscosity and Magnetic Differential Rotation Breaking}

Viscosity efficiently breaks differential rotation~\cite{Kouvaris:2013kra,Markovic:1994bu} and spherical Bondi accretion can be maintained up to BH mass
\begin{equation}
M_{\rm B} = \dfrac{c_s^2}{G}\Big(\dfrac{2\sqrt{3}\nu^2}{\omega_0}\Big)^{1/3} = 1.8 \times 10^{-3}~ (P_1^{1/3}T_5^{-4/3}) ~M_{\odot}~,
\end{equation}
where $\nu = 2 \times 10^{11} T_5^{-2}~\text{cm}^{2}/\text{s}$ is the kinematic sheer  viscosity of the neutron superfluid inside the NS, $\omega_0 = 2 \pi/P$ is the angular velocity, $P_1 = (P / 1\,\text{s})$ is the period and $T_5 = (T/10^5~\text{K})$ is the temperature. On the other hand, the angular momentum of infalling matter can't stall Bondi accretion above a critical BH mass 
\begin{align}
M_{\rm crit} =&~ \dfrac{1}{12^{3/2}}\Big(\dfrac{3}{4 \pi \rho_c}\Big)^2\Big(\dfrac{\omega_0}{G}\Big)^{3} \Big(\dfrac{1}{\psi}\Big)^3\notag\\
=&~ 2 \times 10^{-11}~ P_1^{-3} \Big(\dfrac{\rho_c}{9 \times 10^{14}~\text{g}~\text{cm}^{-3}}\Big)^{-2} \Big(\dfrac{1}{\psi}\Big)^3 ~M_{\odot}~,
\end{align}
where $\psi$ quantifies the BH spin and is 1 for a Schwarzchild BH and 1/3 for an extreme Kerr BH~\cite{Markovic:1994bu}.

Bondi accretion can be violated for central BH masses in the regime $M_B < m_{\rm BH} < M_{\rm crit}$. This could slow down the growth of the central BH substantially.   
The maximal time $t_{\rm viol}$ that a BH can spend violating Bondi accretion is~\cite{Kouvaris:2013kra}
\begin{equation}
t_{\rm viol} = \Big(\dfrac{M_{\rm crit}}{M_{B}}\Big)^2 \tau = 400~ \Big(\dfrac{P}{1\,\text{ms}}\Big)^{-7}\Big(\dfrac{T}{10^7~\text{K}}\Big)^4~\text{yr}~,
\end{equation}
where $\tau = 1/(C_0 M_B)$ and where we have used the values relevant for MSPs:
 $P = 1$\,ms and $T=10^7$ K, where the temperature reflects Bondi accretion heating. Since $t_{\rm viol}$ is short on the relevant time scales ($t_{\rm G}$, $t_{\rm UFD}$), Bondi accretion continues {\it effectively} uninterrupted.

The Bondi-violating mass window $M_B < m_{\rm BH} < M_{\rm crit}$ may be closed 
in the presence of magnetic fields, since magnetic torques can facilitate efficient angular momentum transfer. The initial seed field $B_0$ can be amplified via a dynamo effect.   For the resulting magnetic torque per unit area to exceed the angular momentum current density and ensure rigid rotation, the initial seed field should satisfy the condition~\cite{Markovic:1994bu}
\begin{align}
\dfrac{m_{\rm crit}^{8/3}}{\log m_{\rm crit}} < ~&5.5 \times 10^{10} \Big(\dfrac{x}{0.75}\Big)\Big(\dfrac{6.1 \times10^{2}\,\text{s}^{-1}\,M_{\odot}^{-1}}{C_0}\Big)^2 \notag\\ ~& \times \Big(\dfrac{1\,\text{ms}}{P}\Big) \Big(\dfrac{B_0}{\text{G}}\Big)^2~,
\end{align}
where we take $m_{\rm crit} = M_{\rm crit}/M_B$ and $x\sim 1$ is a parameter describing the fraction of the NS mass that has been consumed.
Here, $C_0$ is a numerical pre-factor of the Bondi accretion as defined in the main text. The resulting condition gives $B_0 > 10^{11}$\,G, which is consistent with magnetic fields associated with pulsars. 

\section{Black hole inside a rotating star, ejected mass}

As the star spins-up, the matter at the equator can exceed the escape velocity and be ejected. For the purpose of making analytic estimates, we take the star's density profile to follow a polytropic relation~\cite{Shapiro:1983du} with index $n = 1.5 - 3$. This range for $n$ subsumes various NS models, from less to more centrally condensed. Some stiffer NS equations-of-state suggest a nearly flat density profiles and, if that is the case, our estimates will be {\it under}-estimates of the amount of ejected material. A polytropic relation connects pressure $P$ and density $\rho$ as $P = K \rho^{(1+1/n)}$, where $K$ is a constant. After a change of variables to $r = \alpha \xi$, where $\alpha$ is a constant, the density is given by $\rho(r) = \rho_c \theta^n (\xi)$, where $\theta(\xi)$ is a solution to the Lane-Emden equation and $\rho_c$ is the central density. The first zero at $\xi_1$ corresponds to the star's radius. For a given mass $M_{\rm NS}$ and radius $R_{\rm NS}$, one obtains $\alpha = R_{\rm NS}/\xi_1$,  
\begin{equation}
\rho_c = M_{\rm NS} / (4 \pi R_{\rm NS}^3) \times |\xi_1 / \theta^{\prime}(\xi_1)|
\end{equation}
with the derivative being with respect to $\xi$.  As the BH inside the NS consumes the star up to radius $r_b = \alpha \xi_b$, the BH mass becomes
\begin{equation}
m_{\rm BH}(r_b) = 4 \pi \alpha^3 \rho_c \xi_b^2 |\theta^{\prime}(\xi_b)|.
\end{equation}
The new NS polar radius $R_p$ is 
\begin{equation}
R_p(r_b) = R_{\rm NS} + R_{\rm S}(r_b) - r_b~,
\end{equation}
where $R_{S}(r_b) = 2 G m_{\rm BH} (r_b)$ is the BH Schwarzchild radius.

For a star rotating near the mass-shedding limit, assuming a
Roche lobe model description for the extended matter envelope (now a rotationally-squashed spheroid), the equatorial radius is related to the polar radius as~\cite{Shapiro:1983du, Shapiro:2002kk} $R_{\rm eq}(r_b) = (3/2) R_p (r_b)$. The equatorial escape velocity is 
$v_{\rm esc} (R_{\rm eq}) = \sqrt{2 G M_{\rm NS} / R_{\rm eq}}$ and is $v_{\rm esc}\sim 0.7$ at the onset of BH growth.
Initially, before the BH size is appreciable, the equatorial velocity is $v_{\rm eq}^0 = \Omega_0 R_{\rm eq}^0 = (3/2) R_{\rm NS} \Omega^0$ for a pulsar rotating at
the initial angular velocity $\Omega_0$.
As the BH grows, the conservation of angular momentum $J = mvr$ for an increment of mass $m$ at the equator determines the equatorial velocity at a later time as $v_{\rm eq}(r_b) = v_{\rm eq}^0 R_{\rm eq}^0 / R_{\rm eq}(r_b)$.
 
Conservation of angular momentum demands that the total initial star angular momentum $J_0$ is distributed between the black hole and the remainder of the neutron star above the Schwarzschild radius:
\begin{align}
J_0 = \dfrac{2}{5} M_{\rm NS} R_{\rm NS}^2 \Omega_0 =& \dfrac{2}{5}(M_{\rm NS} - m_{\rm BH})\Big(\dfrac{R_p^5 - R_{\rm S}^5}{R_p^3 - R_{\rm S}^3}\Big) \Omega \notag\\
&+ J_{\rm BH} + J_{\rm tran}.
\end{align}
Here, the first term on the right describes the angular momentum of the star's solid  spherical shell $(R_p - R_{\rm S})$
in terms of $J_{\rm sh} = I_{\rm sh} \Omega$, where $I_{\rm sh}$ is the shell's moment of inertia.  The last two terms correspond to the angular momentum acquired by the BH itself and the angular momentum transferred out to the remainder of the star. $\Omega (t)$ is the instantaneous angular velocity.
Assuming that the infalling angular momentum is transferred outward efficiently, and that the BH spin contribution is negligible~\cite{Kouvaris:2013kra},
the outer spherical shell acquires an additional angular momentum contribution 
\begin{align}
J_{tran} &= \dfrac{2}{5}(M_{\rm NS} - m_{\rm BH})\Big(\dfrac{R_p^5 - R_{\rm S}^5}{R_p^3 - R_{\rm S}^3}\Big) \Omega_{\rm sh}  \notag\\
&= J_0 - \dfrac{2}{5}(M_{\rm NS} - m_{\rm BH})\Big(\dfrac{R_p^5 - R_{\rm S}^5}{R_p^3 - R_{\rm S}^3}\Big) \Omega~,
\end{align}
where $\Omega_{\rm sh}$ is the additional angular velocity acquired by the outer shell.
For $\Omega (r_b) = v_{\rm eq} (r_b) / R_{\rm eq} (r_b)$, the spin-up yields a new equatorial speed $v_{\rm eq}^{mod} = R_{\rm eq} (\Omega + \Omega_{\rm sh})$. As long as the star is a rigid rotator, 
\begin{equation}
v_{\rm eq}^{mod} = \dfrac{R_{\rm eq} M_{\rm NS} R_{\rm NS}^2}{M_{\rm NS} - m_{\rm BH}} \Big(\dfrac{R_p^3 - R_{\rm S}^3}{R_p^5 - R_{\rm S}^5}\Big) \Omega_0~.
\end{equation}

When the equatorial velocity exceeds the escape velocity, mass is ejected.  For total ejected mass $M_{\rm ej}$, the radius of the rigidly rotating NS is $R_{\rm ej}<R_{\rm NS}$. Hence, the previous equations are modified: 
\begin{align} \label{eq:newejeq}
R_{\rm p,ej} =&~ R_{\rm ej} + R_{\rm S} - r_b ~,\\
R_{\rm eq,ej} =&~ \dfrac{3}{2} R_{\rm p,ej} ~,\\
M_{\rm ej} =& M_{\rm NS} - (4 \pi \alpha^3 \rho_c \xi_{\rm ej}^2 |\theta'(\xi_{\rm ej})|)~, 
\label{Meject}
\\
v_{\rm esc,ej} =&~ \sqrt{\dfrac{2 G (M_{\rm NS} - M_{\rm ej})}{R_{\rm eq,ej}}}~, \\
v_{\rm eq,ej} =&~ \Omega R_{\rm eq, ej}~.
\end{align}
The modified velocity at the new equator after mass ejection, including the central angular momentum transfer,
is $v_{\rm eq, ej}^{\rm mod} = v_{\rm eq}^{\rm mod} (R_{\rm eq, ej} / R_{\rm eq})$.

The condition $v_{\rm esc,ej}( r_b, R_{\rm ej}) \leq v_{\rm eq, ej}^{\rm mod} (r_b, R_{\rm ej})$ signals mass ejection. 
The maximal possible amount of ejected material corresponds to  $R_{\rm ej} = r_b$.
For a pulsar of period $P = 2 \pi / \Omega_0$, we solve for $r_b$ that satisfies 
the above condition with $R_{\rm ej} = r_b$. Having determined $R_{\rm ej}$,
we then numerically integrate the polytropic equation from $R_{\rm ej}$ to $R_{\rm NS}$
to determine the maximal ejected mass $M_{\rm ej}$ from Eq.~(\ref{Meject}). 

While we have neglected BH spin, it is possible for the spin of a slowly-spinning BH to grow to a critical value and thus alter accretion and momentum transfer. However, as found in Ref.~\cite{Kouvaris:2013kra}, this occurs in the regime $M_B < m_{BH} < M_{\rm crit}$ where spherical Bondi accretion is violated. Since such regime is short and occurs during the late stages of BH growth, this consideration will not significantly affect our conclusions.

\section{Allowed Parameter Ranges}

Assuming conservatively that the overall amount of DM should not be larger than the baryonic content, which has a mass of $10^8 M_{\odot}$ in the inner 0.1 kpc volume around the GC, we take the volume-averaged DM density in the CMZ to be $8.8 \times 10^2$ GeV/cm$^{3}$. For velocity dispersion we take~\cite{Kaplinghat:2013xca} 105 km/s.  In a UFD, at 1 pc from the center, the density is~\cite{Navarro:1996gj} $\rho = 1.5 \times 10^{2}$ GeV/cm$^{3}$ and velocity dispersion is~\cite{Koposov:2015jla} $\overline{v} = 2.5$ km/s.

 \begin{figure}[t]
\centering
\hspace{-2em}
\includegraphics[trim = 0.0mm .0mm .0mm .0mm, clip,width = 3.3in]{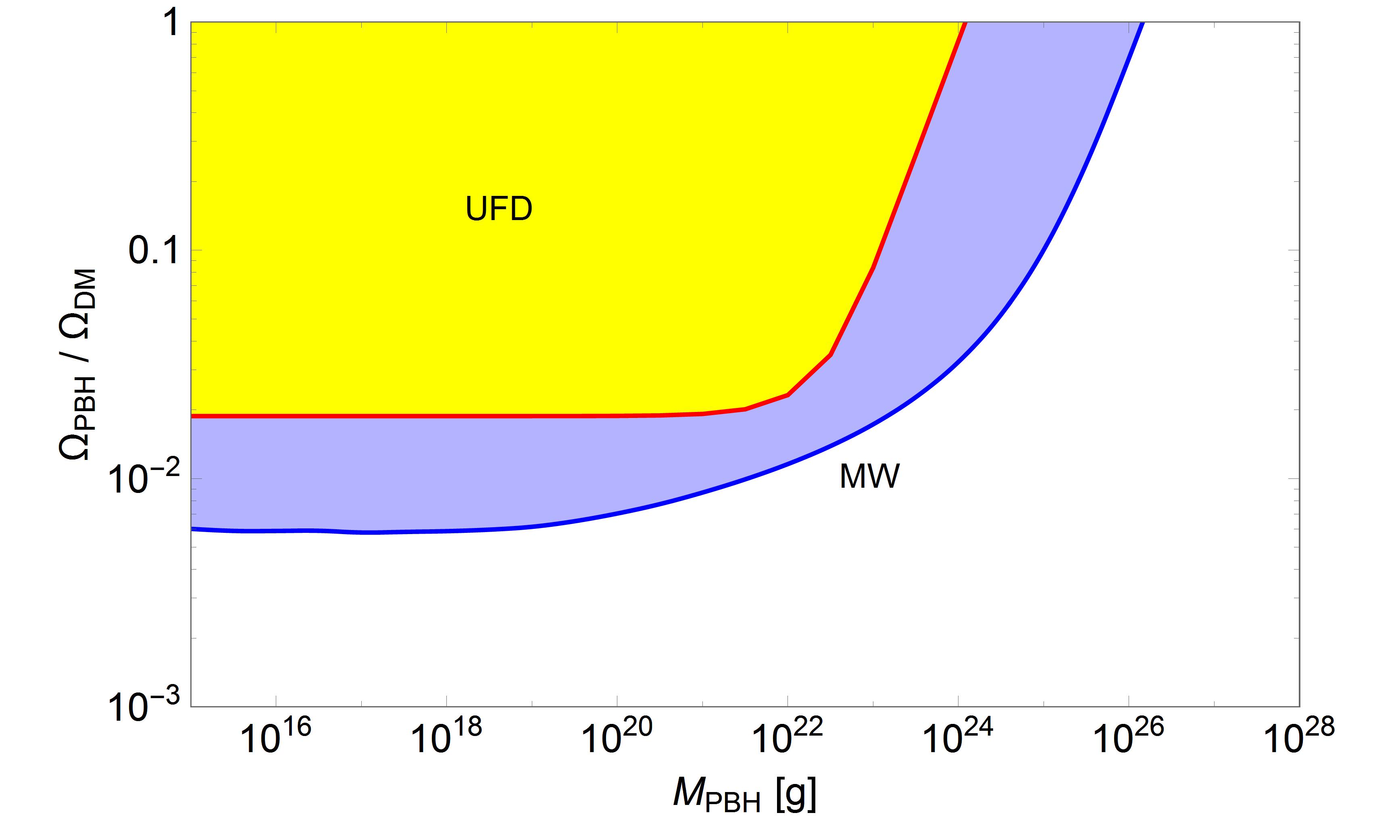}\\
\caption{
Parameter space where PBHs can account for $r$-process element production in the Milky Way and the UFD.}
\label{fig:PBHMWUFD}
\end{figure}
 
To map out the parameter space consistent with the $r$-process abundances of the MW and UFD (Figure~2), we vary the fit input parameters as follows. For the MW, we vary the DM density in the range $50~\text{GeV}/\text{cm}^{3} < \rho_{\text{DM}}^{\text{MW}} < 8.8 \times 10^{2}$ GeV/cm$^{3}$.  The lower bound corresponds to the ``flat-core'' Burkert profile~\cite{Burkert:1995yz} with uniform density in the central kpc.  The upper bound is the volume-averaged maximum allowed mass  of DM within 0.1 kpc of the GC.  For UFD we vary the DM density in the range $0.3~\text{GeV}/\text{cm}^{3} < \rho_{\text{DM}}^{\text{UFD}} < 15$ GeV/cm$^{3}$, which corresponds to the Navarro-Frenk-White~\cite{Navarro:1996gj} (NFW) profile evaluated in the $1 - 50$ pc range from the galactic center, respectively. We consider MW DM velocity dispersion values in the range 
$50~\text{km/s} < \overline{v}^{\text{MW}} < 200~\text{km/s}$, where the lower limit corresponds to possible DM disk within the halo ~\cite{Read:2008fh,Read:2009iv} and the upper limit corresponds to NFW DM density profile without adiabatic contraction~\cite{Kaplinghat:2013xca} at 0.1 kpc from the GC.
Additionally, we take values of the pulsar velocity dispersion in the MW to be in the range of~\cite{Cordes:1997my} 48 km/s to~\cite{Lyne:1998} 80 km/s.
The UFD DM velocity dispersion is varied in the range~\cite{Koposov:2015jla}
$2.7~\text{km/s} < \overline{v}^{\text{UFD}} < 10~\text{km/s}$. Similarly, we also vary the fraction of MSPs not kicked out of the inner $\sim {\rm pc}$ region of the UFD~\cite{Bramante:2016mzo} from 2\% to 25\%. The population-averaged ejected MSP mass is varied in the range $0.09~M_{\odot} < \langle M_{\rm ej} \rangle < 0.32~M_{\odot}$, corresponding to NS density profile that is more centrally  condensed for $n = 3$, or less so for $n = 1.5$ polytropic index, respectively. This provides a conservative estimate of ejected mass, as realistic nuclear equations of state suggest even flatter NS density profiles. In the estimates of ejected mass, we have also included the 1-$\sigma$ uncertainty inherent in the pulsar period-population distribution ~\cite{Cordes:1997my}.
We find that (see main text), within these ranges of parameters, the  UFD and MW $r$-process abundances are explained simultaneously if PBHs contribute at least a few per cent to the 
overall DM density. In Figure~\ref{fig:PBHMWUFD} we display the parameter space span for MW and UFD where PBHs can account $r$-process element production.
 
\vspace{50em}

\end{document}